\documentclass[fleqn,usenatbib]{mnras}

\usepackage[T1]{fontenc}

\DeclareRobustCommand{\VAN}[3]{#2}
\let\VANthebibliography\thebibliography
\def\thebibliography{\DeclareRobustCommand{\VAN}[3]{##3}\VANthebibliography}

\def\equationautorefname#1#2\null{(#2)}

\usepackage{graphicx}	 
\usepackage{amsmath}	 
\usepackage{amssymb}	 
\usepackage{mathtools}   
\usepackage{CJK}         
\usepackage{xcolor}      

\usepackage{scalerel}
\usepackage{tikz}
\usetikzlibrary{svg.path}
\definecolor{orcidlogocol}{HTML}{A6CE39}
\tikzset{
  orcidlogo/.pic={
    \fill[orcidlogocol] svg{M256,128c0,70.7-57.3,128-128,128C57.3,256,0,198.7,0,128C0,57.3,57.3,0,128,0C198.7,0,256,57.3,256,128z};
    \fill[white] svg{M86.3,186.2H70.9V79.1h15.4v48.4V186.2z}
                 svg{M108.9,79.1h41.6c39.6,0,57,28.3,57,53.6c0,27.5-21.5,53.6-56.8,53.6h-41.8V79.1z M124.3,172.4h24.5c34.9,0,42.9-26.5,42.9-39.7c0-21.5-13.7-39.7-43.7-39.7h-23.7V172.4z}
                 svg{M88.7,56.8c0,5.5-4.5,10.1-10.1,10.1c-5.6,0-10.1-4.6-10.1-10.1c0-5.6,4.5-10.1,10.1-10.1C84.2,46.7,88.7,51.3,88.7,56.8z};
  }
}

\usepackage{newtxtext,newtxmath}

\newcommand\orcidicon[1]{\href{https://orcid.org/#1}{\mbox{\scalerel*{
\begin{tikzpicture}[yscale=-1,transform shape]
\pic{orcidlogo};
\end{tikzpicture}s
}{|}}}}

\title[Accretion Column Parameter Survey]{A Parameter Survey of Neutron Star Accretion Column Simulations}

\author[L. Zhang et al.]{
Lizhong Zhang (张力中) \orcidicon{0000-0003-0232-0879},$^{1,2,3}$\thanks{E-mail: lizhong4physics@gmail.com}
Omer Blaes \orcidicon{0000-0002-8082-4573},$^{3}$
Yan-Fei Jiang (姜燕飞) \orcidicon{0000-0002-2624-3399}$^{1}$
\\
$^{1}$Center for Computational Astrophysics, Flatiron Institute, New York, NY 10010, USA\\
$^{2}$School of Natural Sciences, Institute for Advanced Study, Princeton, NJ, USA\\
$^{3}$Department of Physics, University of California, Santa Barbara, CA 93106, USA\\
}

\date{Accepted 2025 June 10. Received 2025 May 20; in original form 2025 March 24}

\pubyear{\the\year{}}

\begin{document}
\begin{CJK*}{UTF8}{gbsn}
\label{firstpage}
\pagerange{\pageref{firstpage}--\pageref{lastpage}}
\maketitle

\begin{abstract}

We conduct a parameter survey of neutron star accretion column simulations by solving the relativistic radiation MHD equations with opacities that account for strong magnetic fields and pair production.  We study how column properties depend on accretion rate, magnetic field strength, and accretion flow geometry.  All the simulated accretion columns exhibit kHz oscillatory behavior, consistent with our previous findings.  We show how the predicted oscillation properties depend on the column parameters.  At higher accretion rates for fixed magnetic field, the column height increases, reducing the local field strength and leading to an anti-correlation between the observed cyclotron line energy and luminosity.  We estimate the line energy from the simulations and find agreement with the observed trend.  Downward scattering in the free-fall zone plays a key role in shaping sideways emission properties and column height.  Strong downward scattering not only re-injects heat back into the column, increasing its height, but also compresses sideways emission, potentially smearing out shock oscillation signals.  When the pair-production regime is reached at the base of the column, the system quickly readjusts to a force balance between gravity and radiative support.  The high opacity in the pair-production region raises the radiation energy density, enhancing sideways emission through a large horizontal gradient.  This shifts the sideways fan-beam radiation toward lower altitudes. In a hollow column geometry, both pencil- and fan-beam radiation emission occurs.  Self-illumination across the hollow region increases the height and stabilizes the inner wall of the column, while shock oscillations persist in the outer regions. 

\end{abstract}




\section{Introduction}

At sufficiently high accretion rates, the magnetically guided accretion flow onto a neutron star is expected to pass through a radiative shock above the neutron star surface, after which a magnetically confined, optically thick accretion column forms \citep{Inoue75,Basko1976}.  In this context, one-dimensional accretion column models typically assume a tophat geometry, while mound-shaped structures are commonly found in two- and three-dimensional calculations \citep{Klein1989,Klein1996,Kawashima2020,Gornostaev2021,Paper2,Paper3,Paper4}.  Models of such accretion columns play an important role in interpreting observed radiation luminosity \citep{Mushtukov2015}, pulse profiles \citep{Klochkov2008,Becker2012}, continuum spectra \citep{Becker07}, cyclotron line spectra \citep{Staubert2019}, and polarization observations \citep{Caiazzo21,Forsblom24} of high luminosity X-ray pulsars in high mass X-ray binaries. 

While often modeled as approximately stationary structures, neutron star accretion columns are almost certainly highly variable due to the presence of entropy waves arising out of the photon-bubble instability in the slow radiative diffusion regime \citep{Arons1992,Klein1996,Hsu1997}. However, detecting that variability has proved challenging \citep{Revnivtsev2015}.  The question then arises as to how far the time-averaged properties of the column match the stationary profiles used in most models.  We have been investigating this question in a series of papers in which we follow the pioneering simulations of \citet{Klein1996} and \citet{Hsu1997} to investigate and characterize the variable structure of neutron star accretion columns.  Using the relativistic radiation MHD code
\textsc{Athena++} \citep{Stone2008}, we first simulated the development of photon bubble instability in a radiation pressure supported magnetized atmosphere \citep{Paper1}, replicating the results of \citet{Hsu1997}.  We then conducted simulations of actual accretion columns, with increasing realism.  We first simulated a column in a vertical magnetic field in Cartesian geometry \citep{Paper2}, then in a split-monopole magnetic field \citep{Paper3}, and then improved the opacities by using classical magnetic scattering rather than non-magnetic Thomson scattering \citep{Paper4}.  In this paper, we present a final suite of numerical simulations with several algorithm improvements, and with further improved opacities \citep{Suleimanov2022}.  We explore a range of accretion rates and magnetic field strengths, include a simulation of a hollow column as well as solid columns, and include a simulation in which pair production produces a dramatic increase in opacity within the column.

\begin{figure*}
    \centering
    \includegraphics[width=\textwidth]{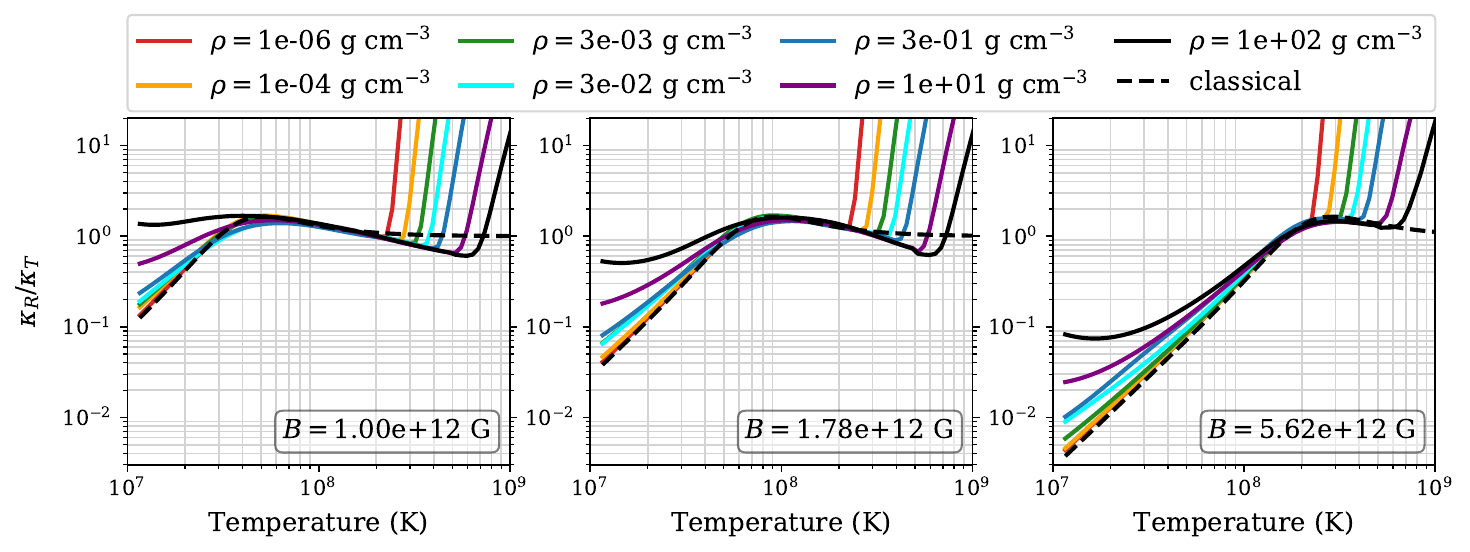}
    \caption{
    Polarization-averaged Rosseland mean opacities of strongly magnetized plasma from \citet{Suleimanov2022}, plotted as a function of temperature for different magnetic field strengths and densities.   Each panel corresponds to a specific magnetic field strength as indicated in the lower-right text box, increasing from left to right.  Variations in density are represented by different line colors, while the black dashed lines indicate the classical approximation, which includes only magnetic scattering and was used in our previous work \citep{Paper4}. 
    }
    \label{fig:opacity}
\end{figure*}

This paper is organized as follows.  In \autoref{sec:Method}, we describe the new algorithm improvements and the simulation setups.  We describe our results in detail in \autoref{sec:Results}, including trends with varying accretion rate and magnetic field (\autoref{sec:vary_mag_and_acc}), the effects of pair production (\autoref{sec:pair_effects}), the changes induced in a hollow column geometry (\autoref{sec:hollow_column}), and the relationship between oscillations of the accretion shock and variations in the output luminosity (\autoref{sec:correlate_shock_and_lum_osc}).  We discuss the observational implications of our results in \autoref{sec:Discussion}, and summarize our conclusions in \autoref{sec:Conclusions}.


\section{Numerical Method}
\label{sec:Method}

Our accretion column simulations are performed in spherical polar coordinates by solving the relativistic MHD equations coupled with the radiation transfer equation, where the radiation intensity is computed in a frequency-integrated, angle-dependent manner.  The governing equations are provided in Section~2.1 of \citet{Paper3}.  The numerical algorithms are described in \citet{Beckwith2011} and \citet{Stone2020} for the relativistic MHD module, \citet{Jiang2014} and \citet{Jiang2021} for the module solving angle-dependent radiation transfer in spherical polar coordinates, and \citet{Paper1} for the coupling of these two modules in \textsc{Athena++} and the necessary modifications to the gravitational source terms.

\subsection{Physics and Algorithm Improvements}

In our previous work investigating the effects of magnetic opacities \citep{Paper4}, we employed a classical treatment of electron scattering in a magnetic field.  We improve on this here by adopting polarization-averaged Rosseland and Planck mean opacities from \citet{Suleimanov2022}, which account for magnetic electron scattering, magnetic bremsstrahlung, cyclotron resonance, vacuum polarization, and electron-positron pairs.  \autoref{fig:opacity} presents the Rosseland mean opacity as a function of magnetic field strength, gas density, and temperature, where $\kappa_R$ and $\kappa_T$ denote Rosseland mean and Thomson opacities, respectively.  The magnetic field strength increases from the left to the right panels, with different line colors representing various gas density values. Dashed lines indicate the classical approximation, which only includes magnetic scattering and is calculated as described in Appendix~A of \citet{Paper4}.  Compared to the classical approximation, the full magnetic opacities increase in the low-temperature and high-density regime due to magnetic bremsstrahlung, and rise sharply beyond the cyclotron peak at high temperatures due to pair creation.  

\textsc{Athena++} evolves the conservative form of the relativistic MHD equations in terms of the conservative variables: mass, total (including magnetic field) momentum, and total energy densities.  In order to carry this out, the normal, primitive variables of the gas, in particular the pressure, are computed from the conservative variables using a variable inversion algorithm.  In a neutron star accretion column, the huge magnetic field strongly constrains the gas motion, and magnetic energy density is orders of magnitude larger than the gas pressure.  As a result, truncation errors from magnetic field reconstruction at the cell center (where fluid pressure is defined) can propagate into the gas thermal energy.  This can lead to unresolved gas temperatures and high-temperature noise, particularly in the free-fall zone where the gas density is low.  Such high-temperature noise has been observed in all of our previous work (e.g., \citealt{Paper2,Paper3,Paper4}).  To mitigate this issue, we implement a backup routine that evolves the entropy equation to compute the gas temperature directly from the first law of thermodynamics.  The gas temperature is corrected using this routine once it is identified as unphysical.  Unlike systems such as accretion disks, the accretion column experiences almost no magnetic reconnection, allowing the entropy equation to provide a complete prescription of energy evolution except near shock regions.  The entropy intrinsically lacks a smooth transition in shock regions and cannot be accurately solved by the Riemann solver.  Therefore, this correction is not applied near shocks.  Details of the relativistic entropy equation and its inversion algorithm are provided in Appendix~C of \citet{LZThesis}.

\subsection{Simulation Setup}

\begin{table*}
\centering
\begin{tabular}{c c c c c c c c c c}
    \hline
    Name & $B_{\star}$ & $\rho_{\mathrm{acc}}$ & $\dot{M}_{\mathrm{acc}}$ & $\left<L\right>_t$ & $\left<L\right>_{\mathrm{iso}}$ & $\eta_{\mathrm{rad}}$ & $f_{\mathrm{peak}}$ \\
    & $\big(10^{12}~\mathrm{G}\big)$ & $\big(10^{-3}~\mathrm{g~cm^{-3}}\big)$ & $\big(10^{17}~\mathrm{g~s^{-1}}\big)$ & $\big(10^{37}~\mathrm{erg~s^{-1}}\big)$ & $\big(10^{39}~\mathrm{erg~s^{-1}}\big)$ & (\%) & (kHz)\\
    \hline
    Mag1-Acc1-S & 1 & \,\,\,2.30 & \,\,\,1.25 & 1.71  & 2.49 & 73 & 6.77 \\
    Mag2-Acc1-S & 2 & \,\,\,2.30 & \,\,\,1.25 & 1.42  & 2.36 & 61 & 7.18 \\
    Mag6-Acc1-S & 6 & \,\,\,2.30 & \,\,\,1.25 & 0.98  & 2.50 & 42 & 8.62 \\
    Mag1-Acc4-S & 1 & \,\,\,9.20 & \,\,\,5.01 & 5.02  & 5.49 & 54 & 3.63 \\
    Mag1-Acc8-S & 1 &      18.40 &      10.00 & 10.84 & 7.74 & 58 & 1.27 \\
    Mag1-Acc1-P & 1 & \,\,\,2.30 & \,\,\,1.25 & 2.22  & 2.65 & 95 & 6.09 \\
    Mag1-Acc2-H & 1 & \,\,\,2.30 & \,\,\,2.51 & 3.11  & 2.37 & 67 & 7.56 \\
    \hline
\end{tabular}
\caption{
Parameters and key properties of the accretion column simulations in this paper.  From left to right, the columns indicate the simulation name, the stellar surface magnetic field strength $B_{\star}$, the gas density $\rho_{\mathrm{acc}}$ at the top of the simulation domain, the mass accretion rate $\dot{M}_{\mathrm{acc}}$, the time-averaged luminosity $\left<L\right>_t$, the time-averaged apparent luminosity $\left<L\right>_{\mathrm{iso}}$ assuming an isotropic radiating source, the radiation efficiency $\eta_{\mathrm{rad}}$, and the frequency $f_{\mathrm{peak}}$ of luminosity oscillation at peak power.
}
\label{tab:sim_param}
\end{table*}

We use two axisymmetric, spherical polar coordinate $(r,\theta)$ grid configurations designed for the solid and hollow geometries of the accretion columns.  In each case, the simulation domain is partitioned into three parts: the accretion column, the vacuum region, and a gas pressure supported base.  In particular, the gas-supported base beneath the accretion column is initialized in hydrostatic equilibrium, with a neutron star surface density of $10^6~\mathrm{g~cm^{-3}}$, and serves as a mass sink that allows accreting material to gradually settle through the sinking zone of the column.  The vacuum region is located at the sides to maintain lateral magnetic confinement of the gas while allowing radiation to escape freely.  The domain partition for the solid column is illustrated in Figure~1 of \citet{Paper3}, with further details and rationale provided in Section~2.2 of \citet{Paper3}.  

The solid accretion column adopts the setup from the fiducial simulation in \citet{Paper3}, which is summarized as follows.  In units of stellar radius $R_{\star}$, the computational domain extends to a height of $1.5R_{\star}$ with an angular width of $0.03$ radians.  Below the neutron star surface ($r < R_{\star}$), the gas pressure supported base has a height of $0.011R_{\star}$, equivalent to approximately 6 pressure scale heights.  The vacuum region surrounds the accretion column above the stellar surface, with an angular width of $0.003$ radians.  The simulation domain is discretized into 7040 cells in the radial direction and 256 cells in the polar direction, with comparable cell widths in both directions, providing a resolution of $129~\mathrm{cm/cell}$ at the base and $323~\mathrm{cm/cell}$ at the top. 

The hollow column follows a similar setup but includes an additional vacuum region near the pole. The height and angular size of the computational domain, including the gas pressure supported base, are identical to those in the solid column setup.  However, the inner edge of the accretion column begins at $\theta = 0.015$, increasing its volume approximately by a factor of 2 compared to the solid column.  Two vacuum regions are positioned inside and outside the accretion column above the stellar surface, with angular widths of $0.015$ and $0.003$ radians, respectively.  The simulation domain consists of 7296 cells in the radial direction and 384 cells in the polar direction, providing a resolution of $125~\mathrm{cm/cell}$ at the base and $313~\mathrm{cm/cell}$ at the top. 

The simulations are initialized using the 1D stationary model of \citet{Basko1976} with a split-monople magnetic configuration, where the magnetic field is specified by the surface field strength and decreases with height as $\propto r^{-2}$.  Similar to the approach in \citet{Paper4}, the magnetic fields used for evolving the MHD equations are initialized to $10^{11}~\mathrm{G}$ at the neutron star surface to provide lateral magnetic confinement.  However, different magnetic field strengths are used to compute magnetic opacity and investigate its impact on the accreting system.  At the top boundary, gas is injected with a density corresponding to different accretion rates.  The sides and bottom of the simulation domain are set to be reflective for gas, in order to stabilize both the surrounding vacuum region and the gas-supported base.  Radiation is allowed to escape freely from the top and sides of the domain.  These boundary conditions are nearly identical to those used in \citet{Paper3}, except for the angular inner boundary, which is set to be reflective to allow higher-order reconstruction and improve numerical performance.  Further details and rationale regarding the boundary conditions can be found in Section~2.2 of \citet{Paper3}. 

We conduct a parameter study with a total of 7 simulations, as listed in \autoref{tab:sim_param}, to explore the effects of magnetic opacities, accretion rates, and columnar geometries on the accretion column structure.  The simulation names indicate three key parameters: the accretion rate ('Acc'), the magnetic field strength used for opacity calculations ('Mag'), and the column type, where 'S' denotes a solid accretion column, 'H' denotes a hollow column, and 'P' denotes a column in the pair production regime.  The surface magnetic field strength $B_{\star}$ in units of $10^{12}$~G is used for computing opacity.  The gas density of the free-fall flow $\rho_{\mathrm{acc}}$ and its corresponding total accretion rate $\dot{M}_{\mathrm{acc}}$ are also listed in cgs units.  Four simulations are performed from the beginning at $t=0~\mathrm{s}$: Mag1-Acc1-S (fiducial), Mag2-Acc1-S, Mag6-Acc1-S, and Mag1-Acc2-H.  For simulations at higher accretion rates (Mag1-Acc4-S and Mag1-Acc8-S), the accretion rate is gradually increased to the target value over $0.99~\mathrm{ms}$ after the fiducial simulation reaches a quasi-steady state at $t=7.08~\mathrm{ms}$.  Note that the simulation time step is very small, approximately $1.70\times10^{-6}~\mathrm{ms}$, limited by the Alfv\'en speed approaching the speed of light.  Since the accretion column at the highest accretion rate (Mag1-Acc8-S) does not naturally enter the pair-production regime, we artificially lower the temperature threshold by using a magnetic opacity table with a fixed density of  $\rho=3\times10^{-3}~\mathrm{g~cm^{-3}}$ (green lines in \autoref{fig:opacity}), as discussed in detail in \autoref{sec:pair_effects}.  This simulation is relaxed from the fiducial simulation at $t=7.08~\mathrm{ms}$ by gradually increasing the magnetic opacity to the target values within $0.99~\mathrm{ms}$.

In order to maintain numerical stability in the simulations against truncation error noise in the gas temperature, we compute the magnetic opacity in optically thick regions from the radiation temperature rather than the gas temperature.  This is a reasonable approximation as local thermal equilibrium (LTE) generally holds in these regions.  We also impose an opacity cap of $8\kappa_T$. This helps avoid gas temperature noise from randomly triggering very large pair opacities above the shock, which can cause the simulations to crash.  

Note that the frequency-integrated opacity in our calculations assumes LTE. The Rosseland mean opacity we use to evaluate gas-radiation momentum exchange is technically valid only in the optically thick regime, where the diffusion approximation holds.  In our simulations with moderately strong magnetic fields (surface field strength of $10^{12}$~G), the gas density even in the free-fall region is sufficiently high to maintain lateral optical depths above unity, making the Rosseland mean a reasonable approximation.  For stronger magnetic fields (i.e., $2\times10^{12}$~G and $6\times10^{12}$~G), the optical depth in the free-fall region can be significantly reduced, breaking the LTE condition.  In such cases, gas-radiation coupling depends on both local gas properties and the radiation field.  Nonetheless, using the approximation formula~(11) from \citealt{Suleimanov2022}), we find that the magnetic Planck mean opacity in these cases happens to be accidentally close to the magnetic Rosseland mean, ranging from $10^{-3}\kappa_T$ to $10^{-1}\kappa_T$ -- well below the Thomson opacity.  This suggests that, near this regime, the opacity is only weakly frequency-dependent.  As a result, the use of magnetic Rosseland mean opacity in our simulations is unlikely to significantly alter the solution relative to a more realistic treatment. 

\section{Results}
\label{sec:Results}

\subsection{Overall Description}

All accretion column simulations exhibit oscillatory behavior due to instantaneous thermal imbalances between global heating and cooling, consistent with our previous findings \citep{Paper2,Paper3,Paper4}.  Finger-like structures are generated by the entropy waves associated with photon bubbles \citep{Arons1992}, as illustrated in \autoref{fig:profile2d}.  To analyze the properties of each accretion column, we select an epoch after the system reaches a quasi-steady state and apply time averaging. Each averaging epoch covers 5 to 12 oscillation periods, except for Mag1-Acc8-S, which includes only 3 periods due to the longer oscillation period and limited computational resources. 

\autoref{tab:sim_param} summarizes the time-averaged measurements (denoted as $\left<\right>_t$) of key properties of each simulation: the intrinsic luminosity $\left<L\right>_t$, the apparent luminosity $\left<L\right>_{\mathrm{iso}}$ (assuming the observer views the column from the side and then interprets it to be radiating isotropically), the radiation efficiency $\eta_{\mathrm{rad}}$ (the fraction of total accretion power converted into radiation), and the peak oscillation frequency $f_{\mathrm{peak}}$ from the luminosity power spectrum, as in \citealt{Paper3}. 

\begin{figure}
    \centering
    \includegraphics[width=0.9\columnwidth]{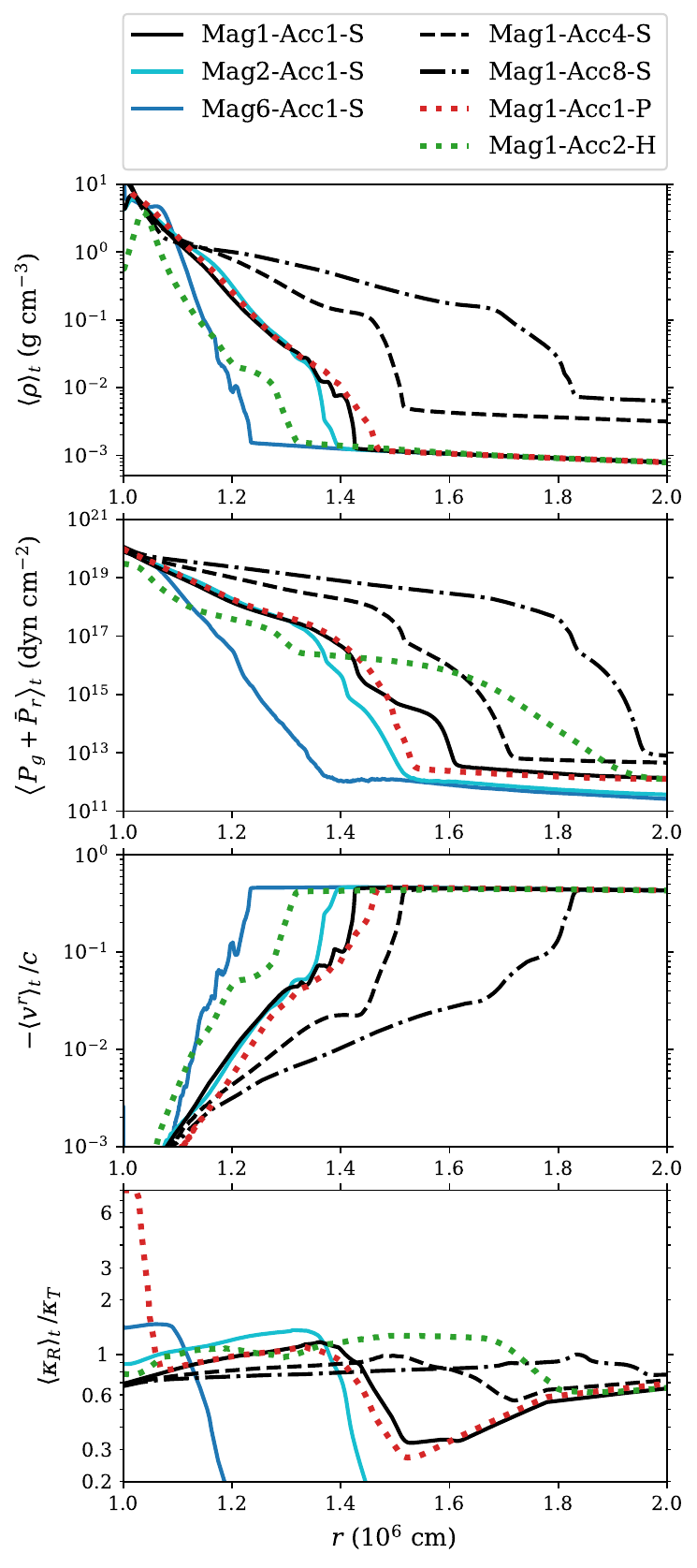}
    \caption{
    Time-averaged 1D radial profiles for all 7 accretion column simulations.  The time averages are taken during the epoch when the accreting systems reach a quasi-steady state.  From top to bottom, the panels display gas density, total thermal pressure, radial fluid velocity, and Rosseland mean opacity, respectively. 
    }
    \label{fig:profile1d}
\end{figure}

\autoref{fig:profile1d} presents time-averaged 1D radial profiles at fixed $\theta$ for all simulations, using different line styles and colors to distinguish various parameters.  The measurements are taken at $\theta=0.01$ for solid columns and $\theta=0.025$ for the hollow column (i.e. a difference of 0.01 in polar angle from the inner side of the column at $\theta=0.015$).   Black lines with varying styles indicate changes in accretion rate, solid lines in multiple colors represent variations in magnetic field strength, and colored dotted lines correspond to the hollow column and pair production runs.  Despite the smoothing effect of time-averaging over radial shock oscillations, the shock structure remains identifiable near regions of sharp changes in density ($\rho$), thermal pressure ($P_g+\bar{P}_r$), and radial velocity ($v^r$).  Here, thermal pressure is defined as the sum of gas pressure and comoving radiation pressure (with the overbar denoting the fluid frame), and $\langle\rangle_t$ indicates a time average.  Below the shock, in the sinking zone, density and thermal pressure exhibit a slow, exponential increase with the sinking velocity gradually decreasing.  The Rosseland mean opacities in the sinking zone are generally near or above the Thompson opacity.  

\begin{figure}
    \centering
    \includegraphics[width=0.9\columnwidth]{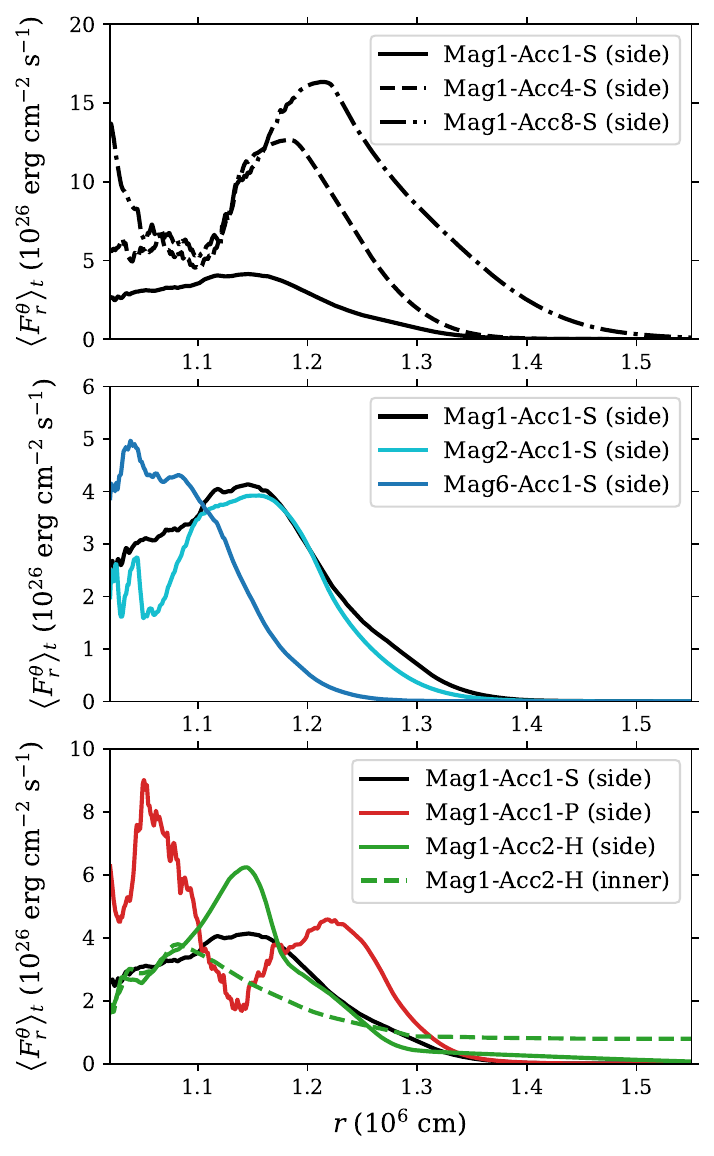}
    \caption{
    Time-averaged sideways radiation emission for all 7 accretion column simulations.  The radiation flux is measured at the sides of the accretion column in the lab frame.  Time averages are taken over epochs when the accreting systems reach a quasi-steady state.  From top to bottom, the panels compare sideways emission across varying accretion rates, magnetic field strengths, and accretion column types, respectively.  Note that the hollow column features vacuum regions at both inner and outer polar boundaries. 
    }
    \label{fig:sideway_flux}
\end{figure}

\autoref{fig:sideway_flux} shows the vertical distribution of time-averaged sideways radiation emission from the vacuum regions, where the sideways radiation flux $F^{\theta}_{r}$ is defined in the lab frame.  The flux peaks indicate the heights at which most radiation escapes from the sides of the accretion column.  These peaks are generally located near or below the shock oscillation regions, except in the pair production run, where the pair production region at the base traps photons and builds up a substantial radiation energy density.  The opacity due to pairs always decreases at the sideways edge of the column because it is cooler there.  This then leads to a large flux leaving the sides due to a large horizontal radiation energy density gradient there.

\begin{figure*}
    \centering
    \includegraphics[width=\textwidth]{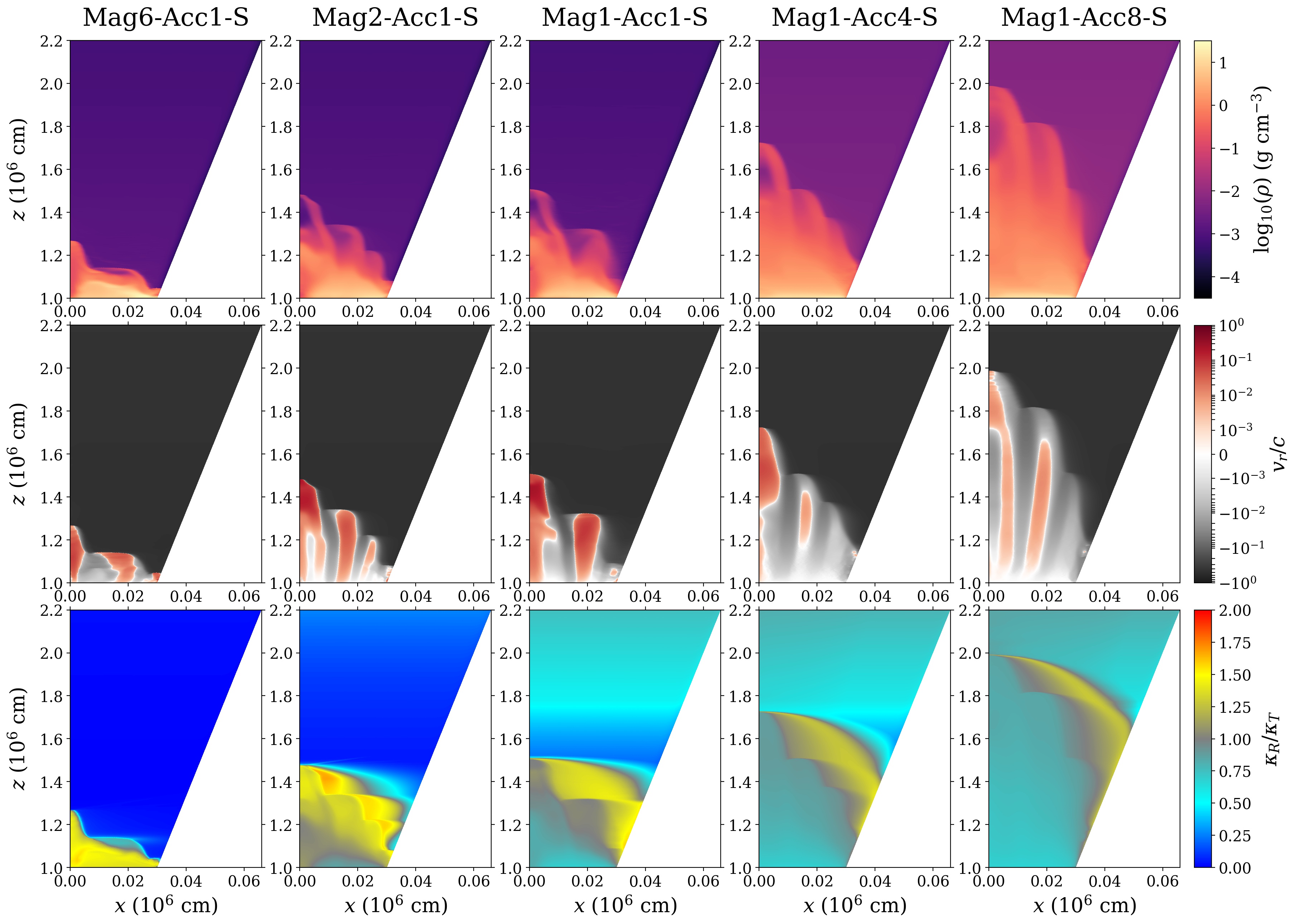}
    \caption{
    Instantaneous 2D profiles of density, radial velocity, and magnetic Rosseland mean opacity at times of maximum radial extent of accretion columns for varying magnetic field strengths and accretion rates.  The middle column represents the fiducial simulation.
    From the first to the middle column, the magnetic field strength decreases, while from the middle to last column, the accretion rate increases.  
    }
    \label{fig:profile2d}
\end{figure*}

In the following sections, we provide a detailed analysis of these simulations.  We compare accretion columns with varying magnetic field strengths and accretion rates (\autoref{sec:vary_mag_and_acc}), investigate the influence of pair production effects on the system (\autoref{sec:pair_effects}), and examine the structure of the hollow accretion column (\autoref{sec:hollow_column}). Finally, we present and explain a correlation between shock oscillations and outgoing radiation (\autoref{sec:correlate_shock_and_lum_osc}). 

\subsection{Varying Accretion Rates and Magnetic Field Strengths}
\label{sec:vary_mag_and_acc}

In order to understand how properties depend on accretion rate and magnetic opacity, we conduct a parameter study of accretion columns by varying the free-fall gas density and surface magnetic field strength.  \autoref{fig:profile2d} presents the instantaneous 2D profiles of accretion columns at their maximum radial extent, showing gas density (top row), radial velocity (middle row), and magnetic opacity (bottom row).  The magnetic field strength decreases at fixed accretion rate from the first to the middle column, while the accretion rate increases at fixed magnetic field strength from the middle to the last.  As shown in the velocity panels, the gas undergoes vertical oscillations below the shock surface, forming finger-shaped structures where adjacent fluid channels move in opposite directions.  These wave patterns arise from entropy waves associated with slow-diffusion photon bubbles, as identified in \citet{Paper3}. 

In the free-fall zone, the gas is cold and its temperature can reach the floor value at high altitudes.  However, as the free-fall gas approaches the shock front and is heated by the emergent radiation, its temperature rises well above the floor values, ensuring that it, and the opacity on which it depends, is physically meaningful.  This is important because, as shown in the density panels of \autoref{fig:profile2d}, a key trend is that the accretion column height increases as the magnetic field strength decreases or the accretion rate increases.  As we now discuss, these height variations are primarily driven by differences in accretion power and downward scattering of radiation above the shock surface.  

That the column height increases with accretion rate is not surprising, as more accretion power is liberated and enhances the thermal pressure within the column.  But there is an additional effect too.  As the accretion rate increases, the denser free-fall flow exerts stronger downward scattering on the emergent radiation from the shock surface.  In particular, radiation from the inner fingers is more strongly beamed sideways due to enhanced downward scattering, which increases the radiation energy density above the shock for the outer fingers.  A fraction of this enhanced radiation energy is further advected down into the accretion column with the free-fall flow, providing additional local heating that increases the column height.  Similarly, when the low-temperature Rosseland mean opacity decreases due to stronger magnetic fields (see \autoref{fig:opacity}), the emergent radiation from the shock surface undergoes less downward scattering.  Consequently, less thermal energy is advected back into the accretion column, resulting in a shorter column structure. 

The downward scattering effect is also evident when comparing the accretion column height (see \autoref{fig:profile2d}) to the vertical distribution of sideways radiation emission (see \autoref{fig:sideway_flux}).  In the strongest magnetic field case (Mag6-Acc1-S), the maximum extent of the accretion column reaches approximately $1.3R_{\star}$, with sideways radiation emission occurring mostly below this height.  In the fiducial run (Mag1-Acc1-S), the accretion column extends up to around $1.5R_{\star}$, but sideways radiation generally escapes below $1.4R_{\star}$.  In the highest accretion rate case (Mag1-Acc8-S), although the accretion column extends to around $2.0R_{\star}$, sideways radiation emission is mostly confined below $1.6R_{\star}$ before escaping the system.  Of course, the time-averaged sideways emission height of an oscillating column (\autoref{fig:sideway_flux}) will always be below the maximum instantaneous height (\autoref{fig:profile2d}).  However, as we discuss in more detail below in \autoref{sec:correlate_shock_and_lum_osc}, even the instantaneous sideways emission height is well below the column height due to downward scattering effects in the free fall zone.  This also influences the transmission of oscillation signals from the shock to the final luminosity.

Below the shock, the density, pressure, temperature, and velocity profiles generally follow classical power-law trends, increasing or decreasing towards the stellar surface.  As shown in the first three panels of the bottom row of \autoref{fig:profile2d}, the Rosseland mean opacity within the column structure generally increases with stronger magnetic fields at the same accretion rate.
Recall that magnetic opacity surpasses Thompson opacity when the fluid temperature approaches the cyclotron energy.  In the fiducial simulation, the temperature increase due to the shock can immediately push the temperature beyond the cyclotron energy.  Once the cyclotron opacity peak is surpassed, a slight decrease in opacity occurs before entering the pair regime as the temperature continues to rise toward the bottom.  However, as shown in \autoref{fig:opacity}, stronger magnetic fields increase the cyclotron temperature and broaden the cyclotron opacity peak, making it more difficult to overcome.  In Mag2-Acc1-S, the shock increases the temperature near the cyclotron energy, but the cyclotron opacity peak is only fully surpassed when the temperature gets high enough near the bottom of the column, as indicated by the slight decrease in opacity.  In Mag6-Acc1-S, the cyclotron opacity peak is too wide to overcome, and the temperature remains near the cyclotron energy, with opacity exceeding the Thompson opacity.

\subsection{Pair Production}
\label{sec:pair_effects}

\begin{figure}
    \centering
    \includegraphics[width=\columnwidth]{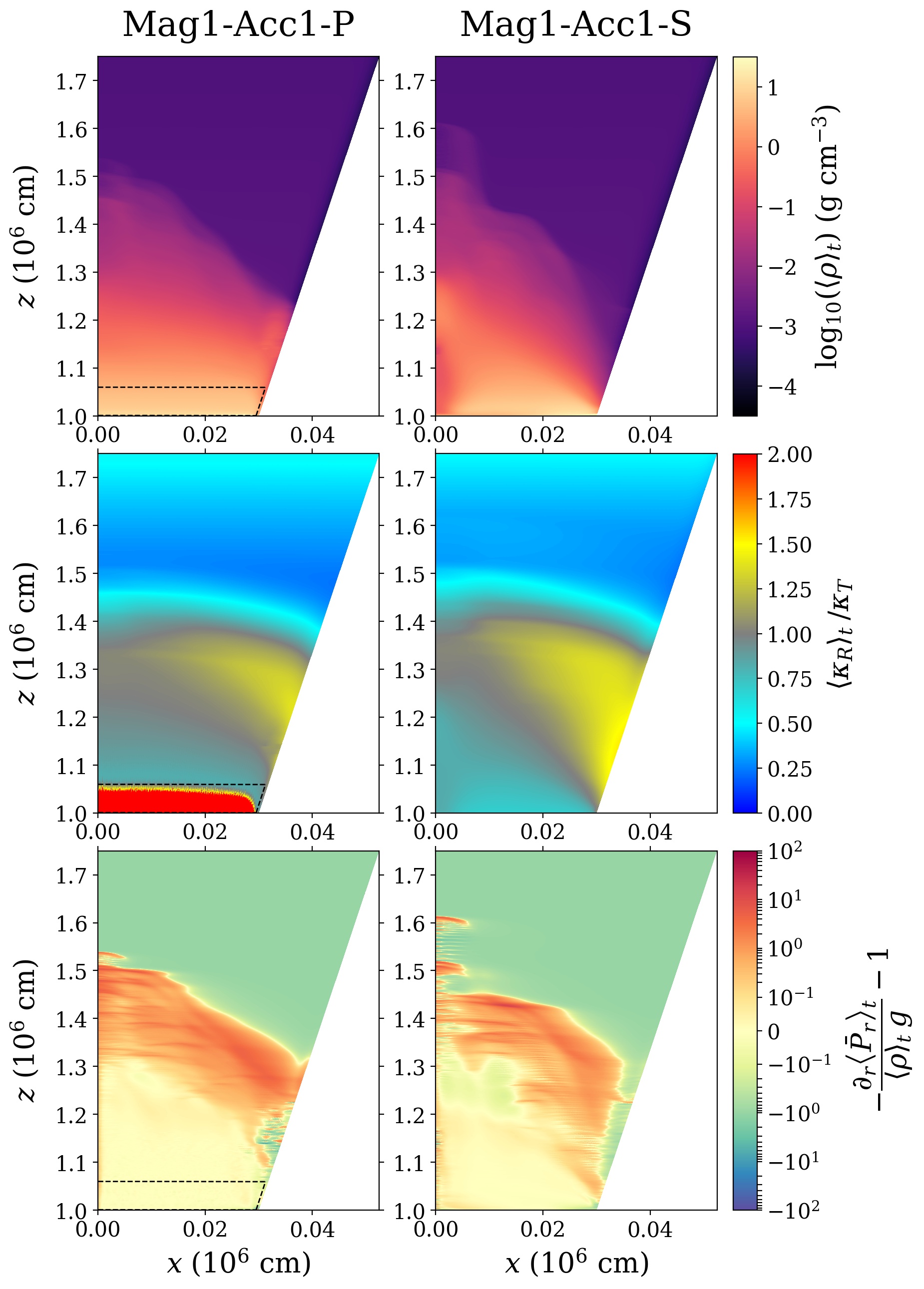}
    \caption{
    Comparison of time-averaged 2D profiles between Mag1-Acc1-P (pair-production run) and Mag1-Acc1-S (fiducial run).  From top to bottom, the panels display density, Rosseland mean opacity, and force balance comparisons.  The black dashed-line box in the left (Mag1-Acc1-P) panels delineates the pair-production region. 
    }
    \label{fig:profile2d_pair}
\end{figure}

Our two runs with increased accretion rates (i.e., Mag1-Acc4-S and Mag1-Acc8-S) were originally designed to reach the pair production regime.  However, neither run achieved this for the following reasons: First, as the accretion rate increases, the density at the base increases slightly, thereby increasing the pair production temperature threshold (see \autoref{fig:opacity}).  Second, when the gas temperature exceeds the cyclotron energy, there is a decline in opacity before reaching the pair production temperature.  This opacity reduction enhances the cooling rate, making it more difficult for the temperature to increase to the pair regime.  We therefore adopted an artificial method to fix the pair production temperature threshold by neglecting the density dependence in the magnetic opacity.  Specifically, we used the magnetic opacity table corresponding to a density of $3\times10^{-3}~\mathrm{g~cm^{-3}}$ (i.e., the green lines in \autoref{fig:opacity}) in our simulations and added an opacity ceiling of 8$\kappa_T$ for numerical stability.  In effect, the opacity we used was therefore a function of temperature only.  This approach allows us to explore how the sharp change in opacity caused by pair production might affect the dynamics of the accretion column.  

In Mag1-Acc1-P, the accretion column enters the pair production regime at the base.  The column remains stable throughout the relaxation phase without any significant dynamical changes.  After reaching a new quasi-steady state, non-linear oscillations persist, with finger-shaped wave patterns propagating from the side toward the center.  \autoref{fig:profile2d_pair} presents a side-by-side comparison of the time-averaged profiles between the pair-production run (Mag1-Acc1-P) and the fiducial run (Mag1-Acc1-S).  The pair-production region at the base is highlighted by a dashed-line box in all the panels on the left.  In the first row, the 2D density distribution becomes more horizontally homogeneous in the pair-production run.  The second row shows that the Rosseland mean opacity enters the pair-production regime at the base.  In the third row, we present the radiation pressure force normalized by gravity (with gravitational acceleration denoted by $g$), indicating that the accretion column region in both runs achieves a time-averaged force balance, despite the large increase in opacity in the pair production region.  

\begin{figure}
    \centering
    \includegraphics[width=0.9\columnwidth]{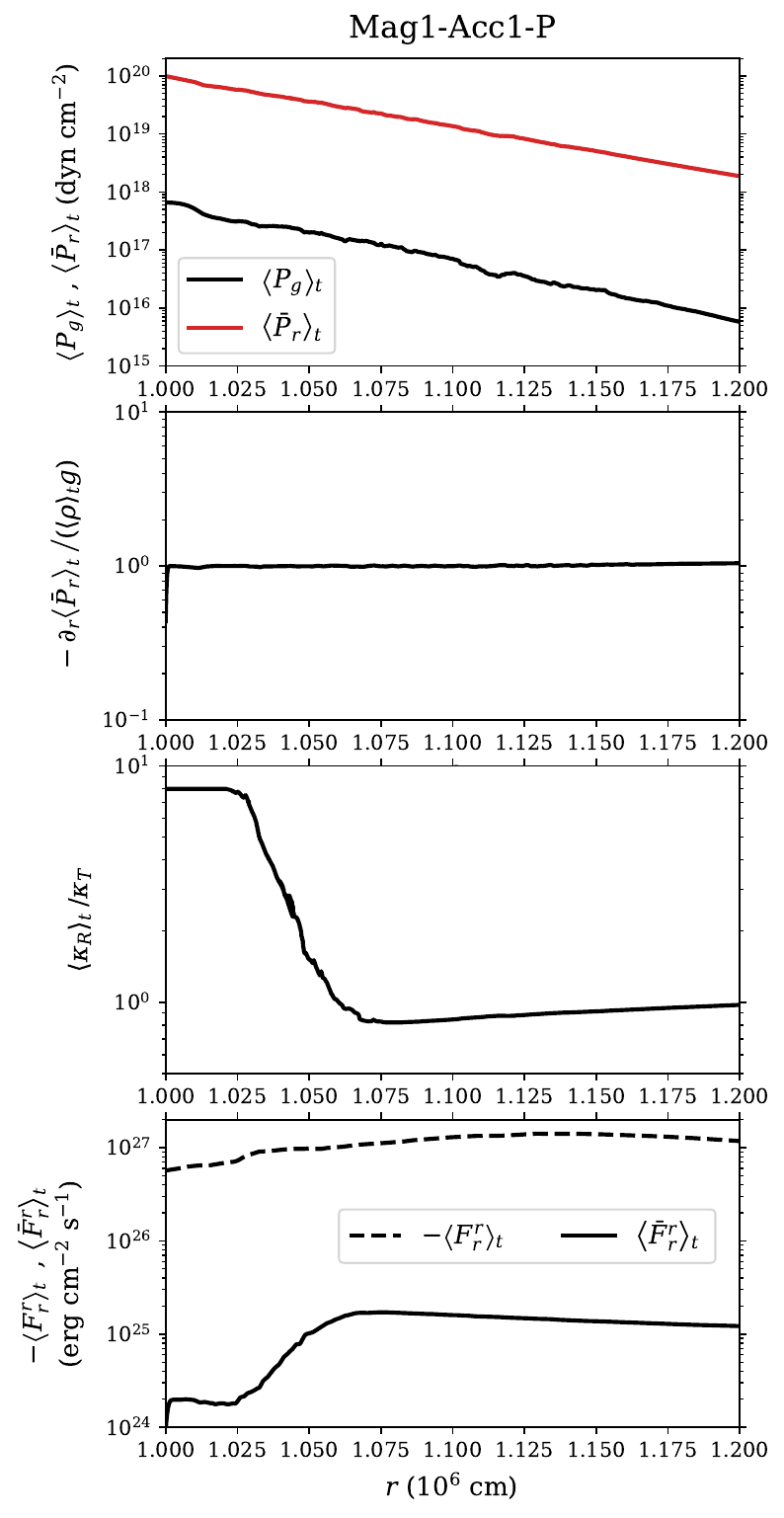}
    \caption{
    Time-averaged 1D profiles of Mag1-Acc1-P (pair-production run) near the base.  From top to bottom, the panels display gas pressure (black) and radiation (red) pressure, the radiation pressure gradient normalized by gravity, Rosseland mean opacity, and radial radiation flux in the fluid frame (solid line) and the lab frame (dashed line).  Note that the lab-frame radiation flux is negative and has been multiplied by -1 for ease of comparison. 
    }
    \label{fig:profile1d_pair}
\end{figure}

To see how this force balance is achieved, we show time-averaged 1D radial profiles near the base at $\theta=0.01$ in \autoref{fig:profile1d_pair}, including pressure, radial pressure gradient, Rosseland mean opacity, and radial radiation flux.  The first panel shows that the accretion column is dominated by radiation pressure, both inside and outside the pair-production region, with gas pressure contributing less than 1\% of the total pressure.  In the second panel, gravity is almost perfectly balanced by the pressure gradient.  The third panel reveals an opacity jump due to pair production, while the radiation flux in the fluid frame drops correspondingly to maintain force balance, as indicated by the solid line in the fourth panel.  Note that the lab-frame radial radiation flux ($F_r^{r}$, dashed line) remains negative (i.e., directed toward the neutron star surface) and is orders of magnitude larger than the fluid-frame radial flux ($\bar{F}_r^{r}$).  This indicates that radiation advection dominates over radiation diffusion.  Force balance is achieved through a small radiation diffusion component in the fluid frame, while energy transport is primarily driven by radiation advection. 

Although force balance is quickly and accurately achieved in the pair production run, thermal equilibrium is not.  Within the pair-production region at the base, as indicated by the dashed-line box in \autoref{fig:profile2d_pair}, heating is dominated by downward radiation advection ($6.83\times10^{36}~\mathrm{erg~s^{-1}}$) and fluid compression through $pdV$ work ($5.55\times10^{36}~\mathrm{erg~s^{-1}}$), while cooling is primarily driven by sideways radiation diffusion ($8.67\times10^{36}~\mathrm{erg~s^{-1}}$).  This results in a net heating rate of $3.71\times10^{36}~\mathrm{erg~s^{-1}}$ within the pair-production region.  Based on this measurement, the temperature in the pair-production region can increase 10\% (50\%) over approximately 20 (170) oscillation periods.  

Given the opacity cap adopted in the simulation ($8\kappa_T$), it would take approximately 500 oscillation periods for the innermost radiation to diffuse out of the pair-production region (i.e., one thermal time).  Therefore, achieving thermal balance in the pair-production region is nearly impossible given limited computational resources, as this would require running the simulation at least 10 times longer.  In reality, the thermal time would be even longer, as realistic pair creation can increase the opacity by orders of magnitude, further hindering the attainment of thermal equilibrium.  However, physically, the temperature increase in this region will eventually stabilize, as the increased photon trapping enhances the sideways radiation energy density gradient, which accelerates sideways radiation diffusion where the opacity drops near the side, helping to restore the thermal balance of the system.  As we noted above, this is evident in excess sideways emission at the bottom in the bottom panel of \autoref{fig:sideway_flux}.  Hence, the enhanced sideways emission resulting from pair creation increases the cooling efficiency of the accretion column and helps prevent it from reaching higher temperatures.  A temperature-limiting effect was also seen in \citet{Mushtukov2019}, though in a higher-temperature and stronger-magnetic-field regime ($> 50$~keV, $>10^{13}$~G) than that explored in our simulations, where the effect is not as prominent.  In that regime, pair production can directly consume a substantial fraction of the accretion power, thereby altering the energy budget of the system and limiting temperature growth.  We do not account for such an effect in our simulations. 

Note that despite the pair-production region absorbing heat, the pair-production run (Mag1-Acc1-P) still shows a 22\% increase in radiation efficiency compared to the fiducial run (Mag1-Acc1-S).  In the fiducial run, about 27\% of the accretion power enters into the gas-supported base (i.e., the neutron star surface), whereas in the pair-production run, only 5\% does.  The pair-production region acts as a new soft bottom boundary, trapping more heat in the column while releasing a portion of it from its side, resulting in an overall increase in radiation efficiency.

\subsection{Hollow Column}
\label{sec:hollow_column}

\begin{figure}
    \centering
    \includegraphics[width=\columnwidth]{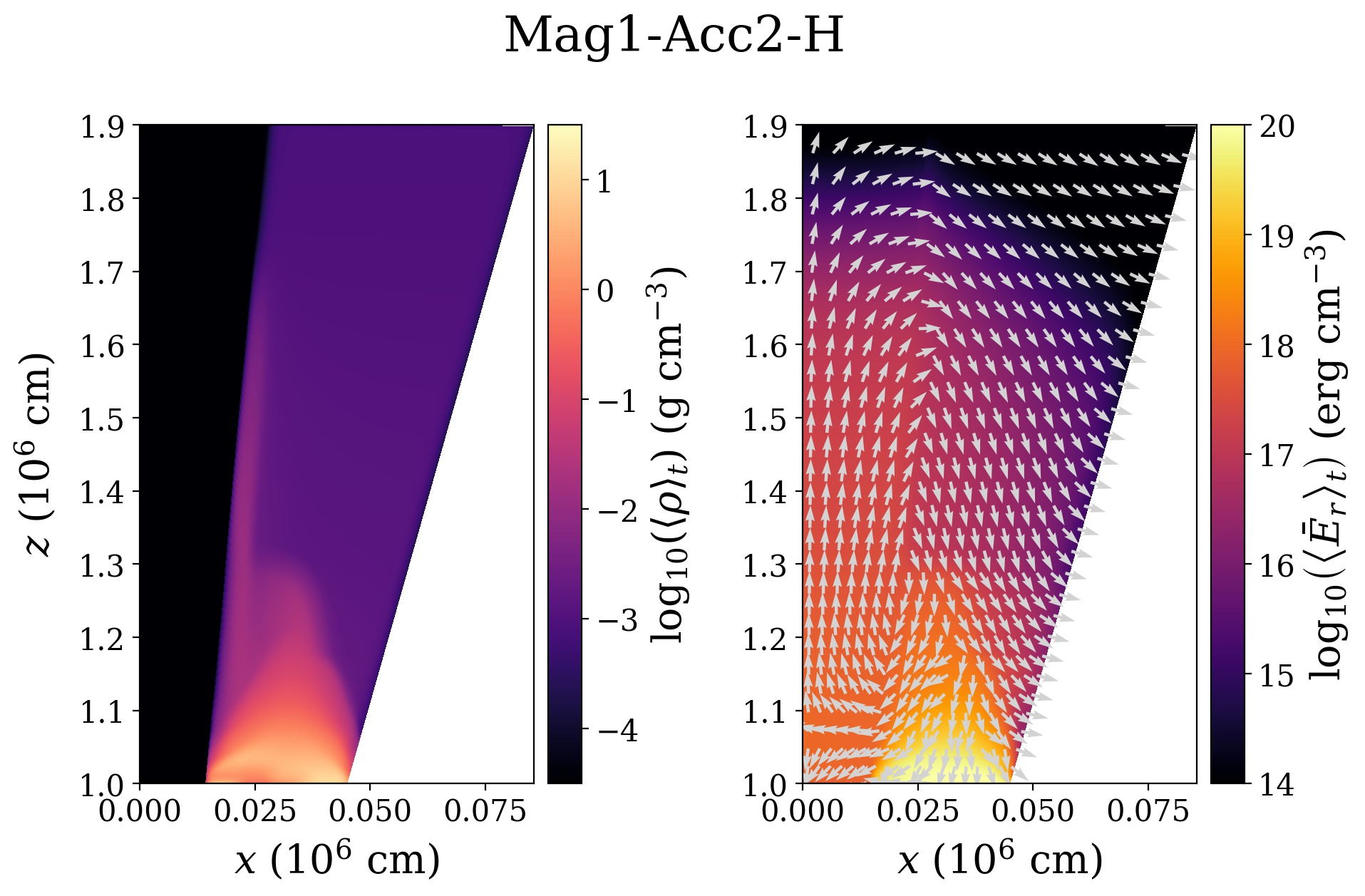}
    \caption{
    Time-averaged 2D profiles of gas density and fluid-frame radiation energy density for the hollow-column run (Mag1-Acc2-H).  The arrows in the right panel indicate the propagation direction of the lab-frame radiation flux. 
    }
    \label{fig:profile2d_hollow}
\end{figure}

\begin{figure*}
    \centering
    \includegraphics[width=0.95\textwidth]{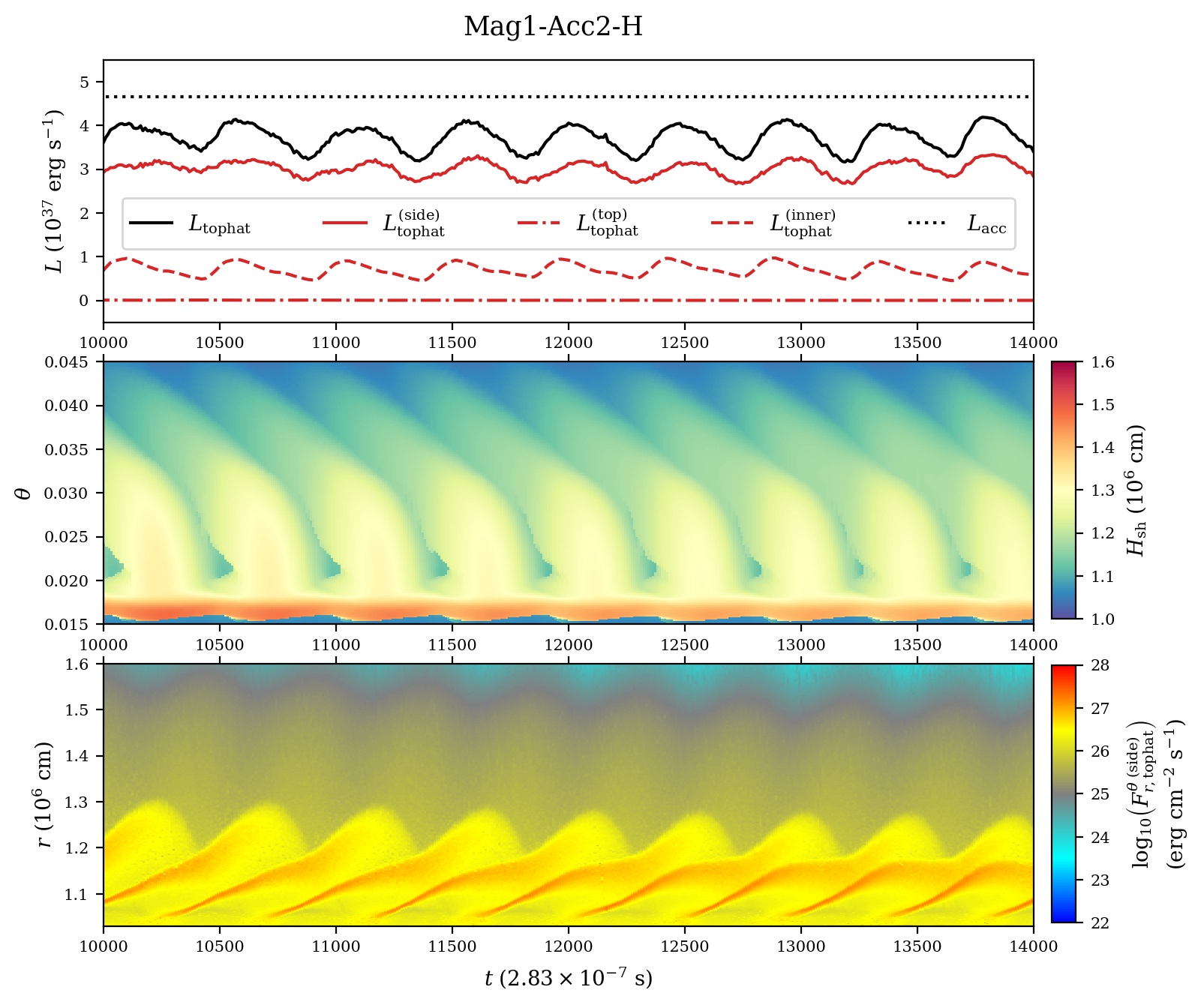}
    \caption{
    Oscillations of shocks and sideways radiation emission in the hollow column simulation.  
    {\bf Top panel:} Time variations in luminosity, showing contributions from radiation emission at the outer side (red solid line), upper side (red dot-dashed line), and inner side (red dashed line).  The total luminosity is represented by the black solid line, while the black dotted line indicates the accretion power.
    {\bf Middle panel:} Space-time diagram of shock oscillations, with color representing the shock height at each time and horizontal distance.  
    {\bf Bottom panel:} Space-time diagram of lab-frame radiation flux at the outer side of the column, with color indicating the magnitude of the sideways flux at each time and height.     
    }
    \label{fig:osc_map_hollow}
\end{figure*}

We simulated one hollow column configuration to investigate its geometric effects on the structure and dynamics of the accretion flow.  Such hollow geometry has also been considered in previous studies, e.g., \citet{Postnov2015}. The setup assumes axisymmetry around the magnetic pole, and is therefore somewhat idealized:  axisymmetry will be a valid assumption only when the angle between the magnetic and spin axes is small and the magnetospheric accretion rate is high.  However, we can still assess how much of our previous results from solid columns still hold under this geometry.  

One major difference compared to the solid column is that, in the hollow column, radiation can directly escape from the sides of hollow region, while at the same time providing self-illumination and heating across the hollow region. In \autoref{fig:profile2d_hollow}, we present time-averaged 2D profiles of gas density and radiation energy density ($\bar{E}_r$) in the fluid frame.  In the hollow column structure, the region closest to the pole extends to higher altitude due to the extra radiative self-illumination, while the region furthest from the pole remains similar to that of the solid column.  

In the right panel of \autoref{fig:profile2d_hollow}, the lab-frame radiation flux is indicated by gray arrows, showing how radiation escapes the system.  Radiation emitted from the inner side back interacts with the column but eventually escapes vertically along the magnetic axis in a pencil-beam pattern. Radiation emission from the outer part of the column remains similar to what is observed in the solid column.  Note that the arrows indicate the direction of the angle-integrated radiation flux and do not directly trace the trajectory of radiation intensity.  Because the hollow region is devoid of plasma, photons fly freely until they escape vertically or intersect the inner column wall or neutron star surface.  The flux arrows therefore indicate the angular distribution of these freely flying photons.  Close to the neutron star the net direction of the photon flux is downward toward the star due to all the radiation emitted from above.  But above $z\simeq1.1\times10^6$~cm, the net flux is outward. 

\begin{figure*}
    \centering
    \includegraphics[width=0.95\textwidth]{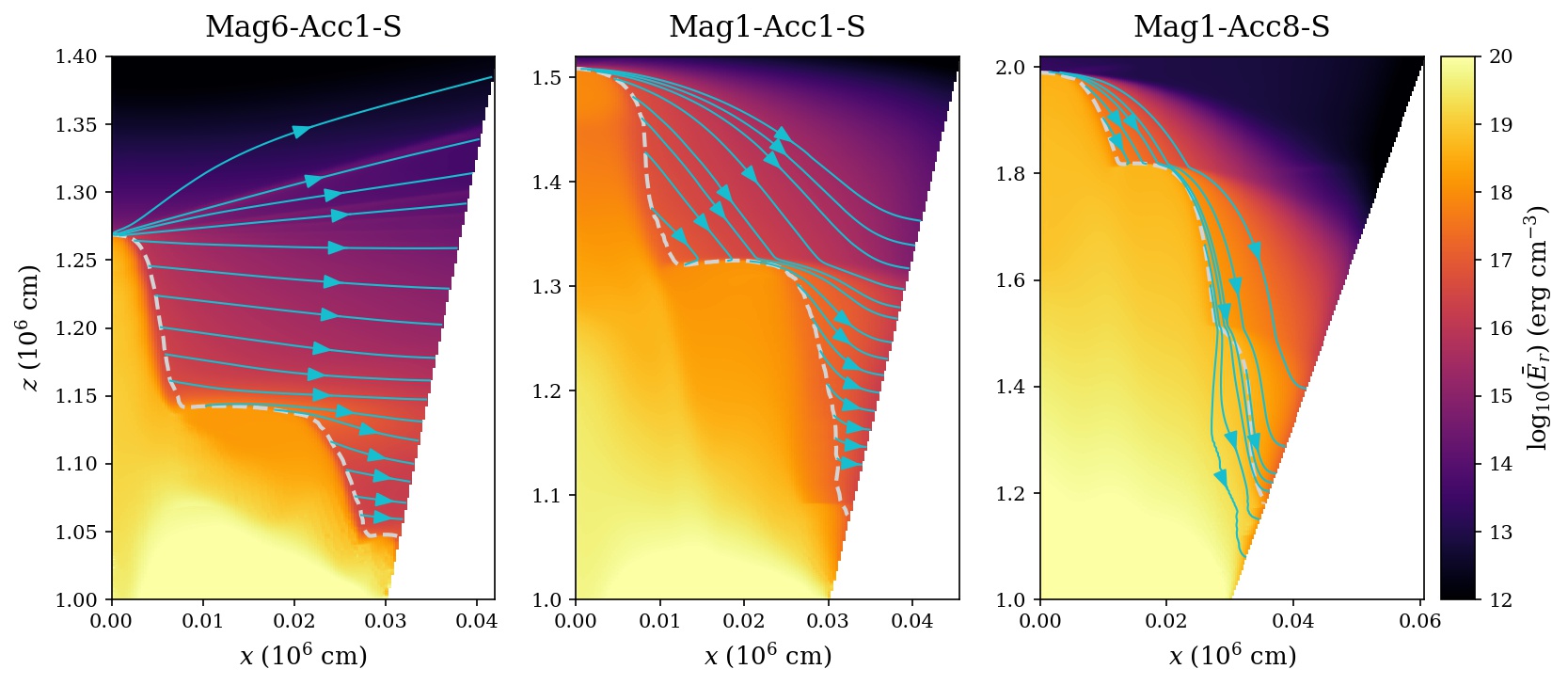}
    \caption{
    Sideways radiation compression effects due to downward scattering.  The 2D fluid-frame radiation energy density profiles are taken from simulation snapshots of Mag6-Acc1-S, Mag1-Acc1-S, and Mag1-Acc8-S at times when the accretion columns are maximally extended.  The gray dashed lines represent the radiation-emergent shock surface.  Blue lines with arrows indicate streamlines of lab frame radiation flux that result in sideways emission from the accretion column. 
    }
    \label{fig:radflux2d}
\end{figure*}

We can evaluate the contributions of inner and outer radiation emission to the total luminosity and examine how shock oscillations influence luminosity variations.  In the upper panel of \autoref{fig:osc_map_hollow}, we measure the time-varying luminosity within a tophat box, defined by $r\le 1.75R_{\star}$ and $0.015\le\theta\le 0.045$.  The total luminosity ($L_{\mathrm{tophat}}$, black solid line) comprises contributions from outer-side emission ($L_{\mathrm{tophat}}^{(\mathrm{side})}$, red solid line), inner-side emission ($L_{\mathrm{tophat}}^{(\mathrm{inner})}$, red dashed line), and top emission ($L_{\mathrm{tophat}}^{(\mathrm{top})}$, red dot-dashed line).  The total accretion power ($L_{\mathrm{acc}}$) is indicated by the black dotted line.  The total luminosity is still dominated by sideways radiation emission, with outer emission accounting for about 75\% and inner emission approximately 25\%.  Emission from the top of the box is negligible.  However, it is important to note that radiation from the inner side escapes in a pencil-beam pattern, while the outer emission forms a fan-beam pattern.  

In the lower two panels of \autoref{fig:osc_map_hollow}, we present space-time diagrams for shock heights ($H_{\mathrm{sh}}$) and the lab-frame sideways radiation flux at the outer boundary of the column ($F_{r,\mathrm{tophat}}^{\theta\ (\mathrm{side})}$).  The oscillation patterns of the shock and sideways emission are well-aligned.  The shock shows minimal oscillation closest to the pole, as indicated by the constant red region near $\theta=0.017$, while most shock oscillations occur beyond $\theta>0.02$.  These shock oscillations are reflected in the overall luminosity variation, with most radiation escaping from the side within the radial range $1.05\le r\le 1.3$, which corresponds to the height of the oscillating shocks ($0.02 \le\theta\le 0.045$).  This behavior is similar to what we found in our previous study of solid accretion columns (see Section~3.3 of \citealt{Paper3} for details). 

However, the strong correlation between shock and luminosity oscillations is observed only when the downward scattering effect in the free-fall region is relatively weak (i.e., when the accretion rate is relatively low or the Rosseland mean opacity is significantly reduced by strong magnetic fields).  As the accretion rate or scattering opacity increases, this correlation is gradually disrupted by the downward scattering process, where compression of the sideways emission region smears out the oscillation signal.  This is further discussed in the next section.

\subsection{Correlation between Shock and Luminosity Oscillations}
\label{sec:correlate_shock_and_lum_osc}

\begin{figure*}
    \centering
    \includegraphics[width=\textwidth]{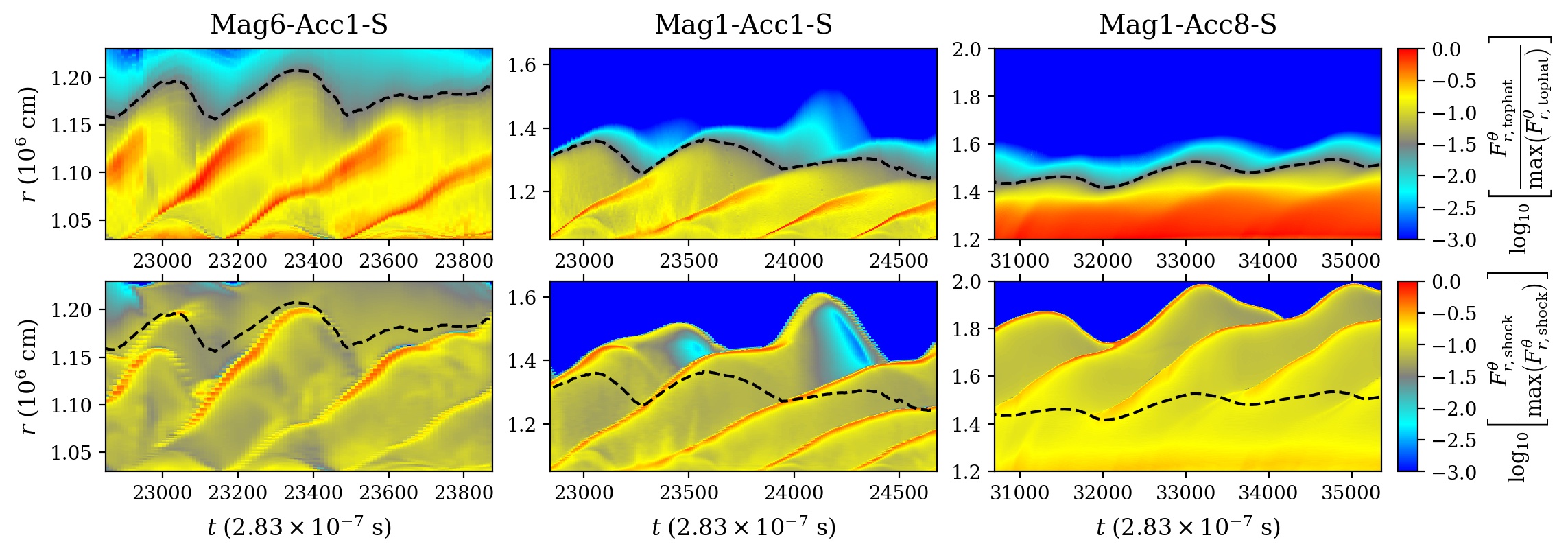}
    \caption{
    Correlation between shock oscillations and luminosity variations.  The upper panels display space-time diagrams of sideways radiation emission from the outer boundary, while the lower panels show space-time diagrams of sideways radiation emerging directly from the shock surface.  The black dashed lines indicate the loci of a particular constant value of sideways radiation flux in the upper panels, and are overplotted in the lower panels for altitude comparison.  
    As downward scattering of radiation becomes stronger from left to right, the compression effects gradually smear out the oscillation signals in the sideways emission (upper panels) that originate from the shock (lower panels). 
    }
    \label{fig:osc_map}
\end{figure*}

In our previous work \citep{Paper3}, coherent oscillatory behaviors were typically found in both shock height and radiation luminosity, which show a strong correlation in the power spectrum (see Figure~8 of \citet{Paper3}).  For a wider and taller column structure, more frequency modes of shock oscillations appear and become imprinted in the sideways radiation emission.  These additional modes can reduce the coherence of radiation pulsations, as seen in Figure~7 of \citet{Paper3}.  These effects primarily occur in cases with relatively low accretion rates, where downward scattering in the free-fall region remains weak. 

However, as the accretion rate increases in Mag1-Acc4-S and Mag1-Acc8-S, we observe a gradual reduction in the luminosity oscillation amplitude.  This is due to enhanced downward scattering, driven by the higher density of the free-fall flow.  In \autoref{fig:radflux2d}, we present simulation snapshots of Mag6-Acc1-S, Mag1-Acc1-S, and Mag1-Acc8-S, illustrating the increasing strength of downward scattering effects from weak to strong.  These snapshots are taken at the moments when the accretion column is maximally extended.  The 2D profiles show radiation energy density in the fluid frame, with gray dashed lines outlining the shock surface and blue lines with arrows indicating the sideways lab-frame radiation flux.  

In the left panel of \autoref{fig:radflux2d}, the strong magnetic field significantly reduces scattering opacity in the free-fall zone, allowing sideways radiation to propagate without noticeable compression effects. In the middle panel, the weaker magnetic field permits downward scattering to emerge, causing radiation emitted from the inner region at an altitude of approximately $1.5R_{\star}$ to be scattered downward to about $1.37R_{\star}$ at the side. In the right panel, the high accretion rate leads to a sufficiently high gas density in the free-fall zone, resulting in strong downward scattering effects. Consequently, radiation emitted from the inner shock is vertically compressed by more than 50\% in height.  Since downward scattering can beam emergent radiation toward the neutron star surface and lead to reflection, we estimate a lower limit on this effect by calculating the fraction of outgoing radiation that intersects the stellar surface (neglecting light-bending effects).  In the fiducial run Mag1-Acc1-S and the hollow-column run Mag1-Acc1-H, approximately 75\% of the output luminosity is intercepted by the stellar surface. For runs with stronger magnetic fields Mag2-Acc1-S and Mag6-Acc1-S, the fractions incident on the stellar surface are 33\% and 3\%, respectively.  In contrast, the high accretion rates cases Mag1-Acc4-S and Mag1-Acc8-S show nearly 100\% of the radiation being reprocessed by the surface.  The pair-production run Mag1-Acc1-P shows an incident fraction of 96\%, as more than half of the beamed radiation escapes near the base of the column. 

Recall that this compression effect occurs in two steps: (1) radiation first undergoes downward scattering and is redirected toward the side, and (2) the increased fluid-frame radiation energy density in the free-fall zone, due to this beaming effect, leads to further downward advection of radiation.

In \autoref{fig:osc_map}, we demonstrate how the downward scattering of radiation affects the transmission of shock oscillations to the final emergent sideways radiation emission.  The upper panels display space-time diagrams of the radiation flux escaping from the outer side boundary ($F_{r,\mathrm{tophat}}^{\theta}$), while the lower panels show space-time diagrams of the sideways radiation flux directly emerging from the shock surface ($F_{r,\mathrm{shock}}^{\theta}$).  The black dashed lines indicate a particular constant value of radiation flux in the sideways emission at the outer boundary and are overplotted in the lower panels for altitude comparisons.  The lower panels effectively track the shock oscillations, as shown by the wave patterns in the space-time diagrams.  In the upper panels, the radiation compression effects increase from left to right, progressively diminishing the oscillation signal of the sideways radiation emission at the boundary.  

In order to observe oscillation signals in the outgoing luminosity, the system must have an appropriate combination of accretion rate and magnetic field strength.  The accretion rate must be sufficiently high to sustain a column structure, but not so high that strong downward scattering suppresses the signal; similarly, the magnetic field must be strong enough to both enable channeled accretion flow and mitigate downward scattering effects.  It is unlikely that such oscillations can be observed in the ULX regime, for two reasons: 1. the accretion rate is typically too high, leading to strong downward scattering of the radiation, and 2. the magnetospheric accretion curtain is likely to be optically thick, causing additional radiation reprocessing and further smearing out the oscillation signal. 

\section{Discussion}
\label{sec:Discussion}

\begin{figure}
    \centering
    \includegraphics[width=0.95\columnwidth]{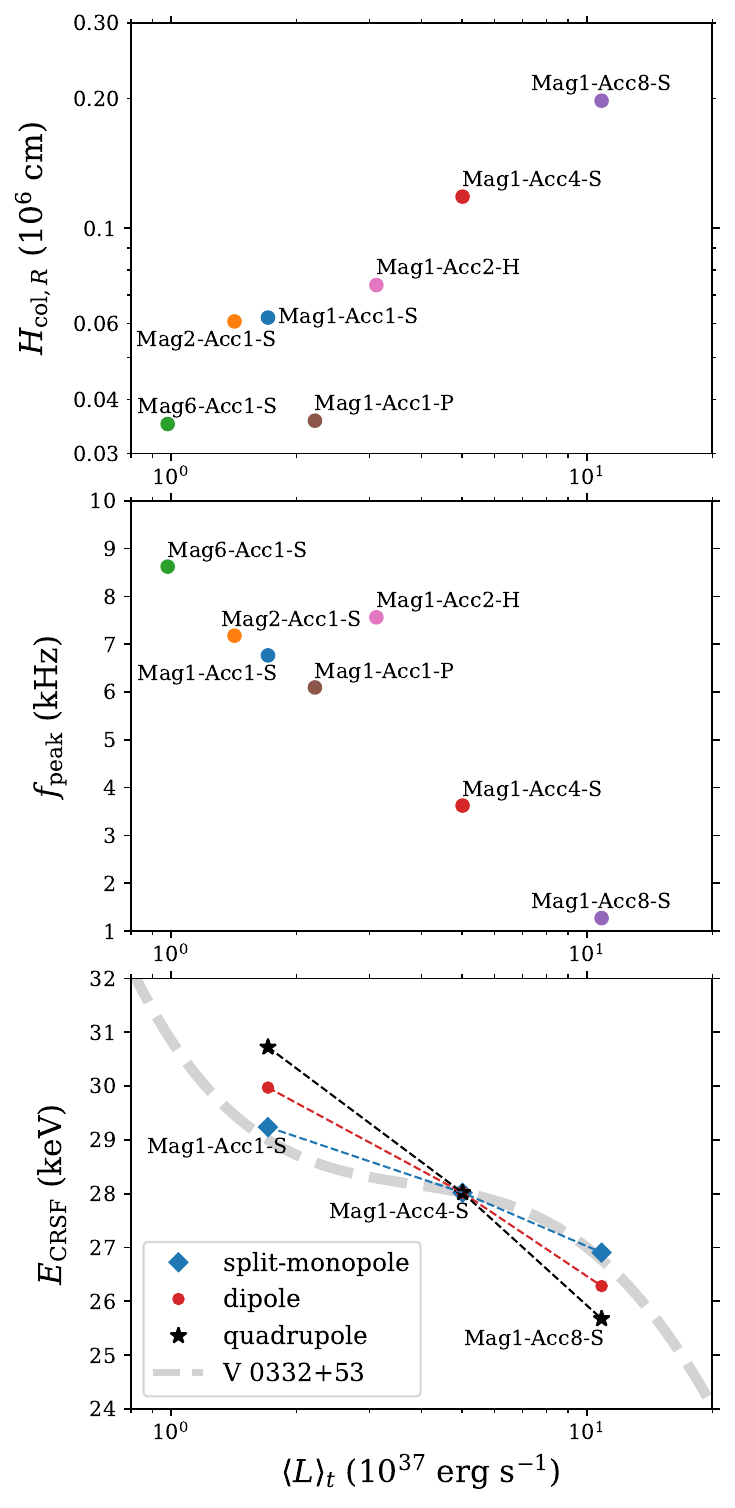}
    \caption{
    Parameter study across all simulations. 
    {\bf Top panel: } Accretion column scale height as a function of luminosity.  The scale height is weighted by the Rosseland mean absorption coefficient ($\rho\kappa_R$).     
    {\bf Middle panel: } Peak oscillation frequency in luminosity as a function of luminosity.  The peak frequency ($f_{\mathrm{peak}}$) corresponds to the frequency with the maximum power in the luminosity power spectrum. 
    {\bf Bottom panel: } Cyclotron line as a function of luminosity.  The dashed line represents the high-order polynomial fit of X-ray pulsar data for source V~0332+53 from \citet{Staubert2019}.  Cyclotron lines from the simulations are inferred from the column height weighted by sideways radiation emission and the rescaled surface magnetic field strength to match observational data near 28 keV, with the gravitational redshift included.  }
    \label{fig:param_study}
\end{figure}

In \autoref{fig:param_study}, we summarize key properties across all simulations that are relevant to observations in this parameter survey.  The top panel presents the scale height of the accretion column weighted by the Rosseland mean absorption coefficients, which better reflects the height of the peak sideways emission (cf. \autoref{fig:sideway_flux}) than density alone.  It is defined as
\begin{equation}
    H_{\mathrm{col},R} = \dfrac{\int_{\mathrm{col}} r \rho\kappa_R dV}{\int_{\mathrm{col}} \rho\kappa_R dV}
    \ ,
    \nonumber
\end{equation}
where the subscript `col' denotes the region within the accretion column.  The accretion column height plays a crucial role in shaping the radiation beaming pattern, potentially leading to different shapes of pulse profiles.  In this plot, the column scale height generally correlates positively with luminosity, making it a useful tracer.  However, in the pair-production run, the formation of a pair-production region significantly increases the Rosseland mean opacity near the column base, resulting in a notable reduction in scale height.  Given that the pair-production regime is inherently difficult to reach, as discussed in \autoref{sec:pair_effects}, this positive correlation can still reliably help estimate accretion column height in most cases.

The middle panel of \autoref{fig:param_study} shows the peak oscillation frequency in the luminosity power spectrum, as in \citealt{Paper3}.  In general, the oscillation frequency is negatively correlated with luminosity, except in the hollow column case.   The reason for this negative correlation is that higher luminosity implies greater accretion heating, which increases the column height, leading to a longer thermal timescale.  In the hollow column case, as discussed in \autoref{sec:hollow_column}, the geometric configuration allows radiation to escape from both the inner and outer sides, thereby enabling both pencil- and fan-beam radiation emission patterns to exist simultaneously.

Searches for high frequency quasiperiodic variability in high luminosity X-ray pulsars have met with mixed success.  Claims of detection \citep{Jernigan2000} were later found to be likely due to instrumental artifacts, and currently only an upper limit
of 0.5\% variability in the frequency range 200-1500~Hz exists for the source V0332+53 \citep{Revnivtsev2015}.  As is evident from the middle panel of \autoref{fig:param_study}, only the very highest luminosity simulation Mag1-Acc8-S has frequencies that low.  However, the downward compression effects in Mag1-Acc8-S likely smear out the oscillation signals in the observed luminosity.  Therefore, to directly observe radiation from shock oscillations in the accretion column system, the accretion rate must be sufficient high to produce super-critical accretion while also remaining low enough -- or with a relatively strong magnetic field -- to mitigate strong downward scattering effects. 

Oscillatory behavior of accretion columns is also reported in \citet{Abolmasov2023}; however, there is a crucial difference between those oscillations and the ones observed in our simulations.  The oscillations in \citet{Abolmasov2023} are likely driven by sound waves and appear stable with damping behavior. In contrast, the oscillations in our simulations arise from nonlinear dynamics triggered by the system's inability to maintain instantaneous thermal balance between heating and cooling within the column \citep{Paper2}.  Because the model used in \citet{Abolmasov2023} adopts a 1D framework, it cannot capture the 2D entropy waves associated with photon bubbles. In our simulations, these 2D entropy modes play a central role in shaping the column structure and enable self-consistent sideways radiative cooling, which sustains the nonlinear oscillations. 

The bottom panel of \autoref{fig:param_study} directly compares the simulation results with observational data.  The dashed line represents the high-order polynomial fit for the source V~0332+53 from \citet{Staubert2019}.  The negative correlation between luminosity and cyclotron resonant scattering features (CRSF) suggests that the system is likely an accretion column, where a taller structure at higher luminosity reduces the local magnetic field strength under non-uniform magnetic geometries such as split-monople, dipole, and quadrupole.  As illustrated in Section~3.4 of \citet{Mushtukov2022} and Section~4 of \citet{Staubert2019}, a broader population of X-ray pulsars also exhibits this negative correlation. 

To evaluate the expected cyclotron lines from our simulations, we first determine the column height weighted by the sideways radiation flux, which is primarily locates at the flux peak, as shown in \autoref{fig:sideway_flux}.  This sideways flux weighting effectively tracks the height associated with emergent radiation, which likely determines the local magnetic field strength for the observed cyclotron line.  Next, we rescale the surface magnetic field strength in a series of runs at varying accretion rates (i.e., Mag1-Acc1-S, Mag1-Acc4-S, and Mag1-Acc8-S) so that the gravitationally redshifted cyclotron line for the middle case (Mag1-Acc4-S) aligns with observational data near 28~keV.  We also interpret our simulation data by considering different magnetic field topologies.  In addition to the split-monopole field ($\propto r^{-2}$, blue diamonds) used in the simulations, we compute the inferred cyclotron lines for dipole ($\propto r^{-3}$, red circles) and quadrupole ($\propto r^{-4}$, black stars) fields.  Note that our estimate of the trend in dipole and quadrupole fields is likely an overestimate, because it does not account for the increased horizontal area of the higher multipole fields with height.  The same mass column would therefore be spread out over a larger area, resulting in a reduction in height that we do not account for.  Moreover, reflection of the emergent radiation from the accretion column off the neutron star surface can further reduce the observed cyclotron energy \citep{Poutanen2013, Lutovinov2015}, particularly when the column is tall and downward scattering is significant, such that a large fraction of the radiation is beamed towards the stellar surface.  A more quantitative evaluation of this negative correlation requires a complete post-processing of the simulation data, as done in \citet{Loudas2024}.  In the future, we plan to perform post-processing on our parameter survey of neutron star accretion column simulations with the neutron star surface included, to establish a more direct connection to observables, such as the spectrum and polarization. 

\section{Conclusions}
\label{sec:Conclusions}

For the first time, we conduct a parameter survey of neutron star accretion column simulations using a prescribed frequency and polarization-averaged opacity that fully accounts for strongly magnetized plasma and pair production effects.  This survey provides a comprehensive physical picture and detailed spatial profiles of neutron star accretion columns as a function of magnetic field strength, accretion rate, and accretion geometry.

All accretion columns exhibit a highly oscillatory nature, where taller columns have slower shock oscillations due to their longer thermal timescales.  At the same accretion rate, a stronger magnetic field results in a shorter column due to weaker downward scattering above the shock.  Conversely, at the same magnetic field strength, higher accretion rate leads to a taller column due to greater energy input.  Since the stellar magnetic field decreases with height, this provides a quantitative explanation for the negative correlation between the observed cyclotron line energy and luminosity in super-critical X-ray pulsars, such as V~0332+53. 

Downward scattering in the free-fall zone plays a crucial role in shaping the accretion column height and sideways radiation emission.  Strong downward scattering can increase column height by partially re-injecting emergent radiation back into the column through downward advection in the free-fall zone.  It also spatially compresses sideways emission, potentially smearing out shock oscillation signals.  Our parameter survey provides a theoretical constraint on detecting shock oscillation signals via radiation emission: the accretion rate must be high enough to sustain a radiation-dominated column, but if the gas-radiation interaction is too strong -- either due to an excessively high accretion rate or a relatively weak magnetic field -- downward scattering becomes dominant, compressing sideways emission and erasing the oscillation signal.

When the pair-production regime is reached at the base of the accretion column, the system quickly readjusts to a new force balance, where gravity is counteracted by the radiation pressure gradient under the diffusion approximation.  In this case, the high opacity in the pair-production region results in a low upward radiation flux in the fluid frame.  However, this does not affect heating within the column, as heating is primarily driven by radiation advection in the lab frame.  Meanwhile, the accretion column dynamics remains unchanged, continuing to exhibit the same regular nonlinear oscillatory behavior as in other cases.  In addition, since the pair-production region acts as a reservoir of trapped photons and provides a large radiation energy density gradient for sideways emission, the vertical distribution of fan-beam radiation shifts toward lower altitudes, altering the emergent radiation beam pattern. 

When the magnetic and rotational axes are nearly aligned, super-critical accretion onto a strongly magnetized neutron star can result in a hollow column geometry.  This structure can generate both pencil- and fan-beam radiation emission patterns.  Although the inner accretion shock is stabilized by additional radiative self-illumination within the hollow region, shock oscillations in the outer regions remain unchanged. 

\section*{Acknowledgements}
We thank Nick Loudas for useful conversations, and the referee for helpful remarks that improve this paper.  This work was initiated by LZ at UCSB, supported by NASA Astrophysics Theory Program grant 80NSSC20K0525.  LZ completed all simulations at IAS with support from Schmidt Futures and NASA, which is part of the TCAN collaboration supported by the grant 80NSSC21K0496.  The analysis was conducted by LZ at the Center for Computational Astrophysics at the Flatiron Institute, which is supported by the Simons Foundation.  Computational resources for this work were provided by the NASA High-End Computing (HEC) Program through the NASA Advanced Supercomputing (NAS) Division at Ames Research Center.  The analysis made significant use of the following packages: NumPy \citep{Harris2020}, SciPy \citep{Virtanen2020}, and matplotlib \citep{Hunter2007}.

\section*{Data Availability}

All the simulation data reported here is available upon request to the authors.



\bibliographystyle{mnras}
\bibliography{references} 







\bsp	
\label{lastpage}
\end{CJK*}
\end{document}